\documentclass[twocolumn,twoside,slac_two]{revtex4}
\usepackage{graphicx}
\usepackage{fancyhdr}
\newcommand{\bzcat}{ROMA-BZCAT}

\newcommand{\fer}{{\it Fermi}}

\newcommand{\swf}{{\it Swift}}

\newcommand{\wse}{{\it WISE}}

\begin{document}
\title{Unveiling the nature of the unidentified gamma-ray sources}

\author{F. Massaro}
\affiliation{SLAC National Laboratory and Kavli Institute for Particle Astrophysics and Cosmology, 2575 Sand Hill Road, Menlo Park, CA 94025, USA}
\author{R. D'Abrusco, A. Paggi, H. A. Smith}
\affiliation{Harvard - Smithsonian Astrophysical Observatory, 60 Garden Street, Cambridge, MA 02138, USA}
\author{N. Masetti}
\affiliation{INAF - Istituto di Astrofisica Spaziale e Fisica Cosmica di Bologna, via Gobetti 101, 40129, Bologna, Italy}
\author{M. Giroletti}
\affiliation{INAF Istituto di Radioastronomia, via Gobetti 101, 40129, Bologna, Italy}
\author{G. Tosti}
\affiliation{Dipartimento di Fisica, Universit\`a degli Studi di Perugia, 06123 Perugia, Italy}

\begin{abstract}
One of the main scientific objectives of the recent \fer\ mission
is investigating the origin of the unidentified gamma-ray sources (UGSs).
Despite the large improvements of Fermi in the gamma-ray source
localization with respect to the past gamma-ray missions,
about 1/3 of the gamma-ray objects detected do not
have yet an assigned counterpart a low energies.
We recently developed a new association method to identify
if there is a $\gamma$-ray blazar candidate within the positional 
uncertainty region of a generic $\gamma$-ray source.
This method is entirely based on the discovery that blazars, 
the largest known class of gamma-ray sources,
can be recognized and separated from other extragalactic
sources on the basis of their infrared colors.
Here we summarize the results obtained by applying 
our association procedure to the 
unidentified $\gamma$-ray sources (UGSs) and to the
active galaxies of uncertain type (AGUs) listed in the Second \fer\ Large Area Telescope (LAT) catalog.  
\end{abstract}

\maketitle
\thispagestyle{fancy}

\section{Intorduction}
Unveiling the nature of the Unidentified Gamma-ray Sources (UGSs) \citep{abdo09}
is one of the biggest challenges in contemporary gamma-ray astronomy.
In particular, according to the Second \fer\ Large Area Telescope (LAT) catalog \citep[2FGL;][]{nolan12}, 
$\sim$1/3 of the $\gamma$-ray detected sources are still unassociated with their low energy counterparts. 
A large fraction of the UGSs are likely to be blazars, the rarest class of active galaxies, because
their emission dominates the $\gamma$-ray sky \citep[e.g.,][]{mukherjee97,abdo10}.

However, due to the incompleteness of the current radio and X-ray surveys on the basis of the usual
$\gamma$-ray association method is not always possible to find the blazar-like counterpart of an UGS.
Additional attempts have also been recently developed to associate or to characterize the UGSs 
using either pointed \swf\ observations \citep[e.g.,][]{mirabal09a,mirabal09b} 
or statistical approaches \citep[e.g.][]{mirabal10,ackermann12}. 

Blazar emission is characterized by high and variable polarization, 
apparent superluminal motions, and high luminosities,
coupled with a flat radio spectrum that steepens toward the infrared-optical bands and
together with rapid flux variability at all frequencies \citep[e.g.,][]{urry95}.
Their broad band spectral energy distributions show two main 
components: the low-energy one with power peaking in the range from the IR to the X-ray band, 
and the high-energy showing its maximum in the MeV -- TeV energy range \citep[e.g.,][]{giommi05}.

Blazars come in two flavors: the BL Lac objects,
characterized by featureless optical spectra and lower luminosity with respect to the second class
composed of flat-spectrum radio quasars showing quasar-like optical spectra \citep{stickel91,stoke91}.
In the following we label the BL Lac objects as BZBs and the 
flat-spectrum radio quasars as BZQs, following the nomenclature 
of the Multifrequency Catalogue of Blazars \citep[\bzcat,][]{massaro09,massaro10,massaro11}.

On the basis of the preliminary data release of the Wide-field Infrared Survey Explorer
\citep[\wse, see][for more details]{wright10}, 
we discovered that in the 3-dimensional IR color space $\gamma$-ray emitting blazars
lie in a distinct region, well separated from other extragalactic 
sources whose IR emission is dominated by thermal radiation 
\citep[e.g.,][]{paper1,paper2}.
According to D'Abrusco et al. (2013) we refer to the 3-dimensional region occupied 
by $\gamma$-ray emitting blazars as the $locus$, 
to its  2-dimensional projection in the [3.4]-[4.6]-[12] $\mu$m color-color diagram 
still maintain its historical definition as the \wse\ Gamma-ray Strip.

This \wse\ analysis led to the development of a new association 
method to recognize $\gamma$-ray blazar candidates
for the unidentified $\gamma$-ray sources listed in the 2FGL
\citep{paper3,paper4}, as well as in the 4$^{th}$ {\it INTEGRAL} catalog \citep{paper5}.

Here we present the results achieved by applying 
a more conservative approach and several improvements recently made on the association procedure 
\citep[see][for more details]{paper6}, mostly based on the availability of the \wse\ full archive
\footnote{http://wise2.ipac.caltech.edu/docs/release/allsky/}, available since March 2012 \citep[see also][]{cutri12}.
We analyzed all the UGSs listed in the 2FGL as well as
the sample of the active galactic nuclei of uncertain type (AGUs) that have still unclear classification \citep{nolan12}.
We also performed an extensive literature search looking for 
multifrequency information on the $\gamma$-ray blazar candidates selected on the basis of their \wse\ colors
to confirm their nature. 

{\it We remark that a detailed description of our multifrequency analysis and of the association procedure adopted, 
together with the lists of $\gamma$-ray blazar candidates,
is presented in details in Massaro et al. (2013) and D'Abrusco et al. (2013), 
while here we only highlight the major results achieved.}

\section{RESULTS}
A new association method has been recently developed based on the discovery that 
$\gamma$-ray emitting blazars lie in a distinct region in the \wse\ 3-dimensional color space,
separated from that occupied by other extragalactic and galactic sources \citep{paper1,paper2}.
According to D'Abrusco et al. (2013) the 3-dimensional region occupied 
by $\gamma$-ray emitting blazars is the $locus$;
its 2-dimensional projection in the [3.4]-[4.6]-[12] $\mu$m parameter space,
retains its historical definition of \wse\ Gamma-ray Strip \citep{paper1}.
Additional improvements, mostly based on the \wse\ all-sky data release, available since March 2012 \citep{cutri12},
and on a new parametrization of the $locus$ in the parameter space 
of its principal components have been subsequently developed \citep{paper6}.

Then we applied our new association procedure
searching for new $\gamma$-ray blazar candidates in the two samples: 
the unidentified gamma-ray sources (UGSs), and the active galaxies 
of uncertain type (AGUs), {\bf as} listed in the 2FGL \citep{nolan12}. 
We also perform an extensive archival search to see if the sources associated with our method,
show additional blazar-like characteristics; as, for example, the presence of a radio counterpart and/or of a spectrum that could be 
featureless as for BZBs or similar to those of broad-line quasars as generally occurs in BZQs.

\begin{figure}[!ht] 
           \includegraphics[height=9cm,width=8.8cm,angle=0]{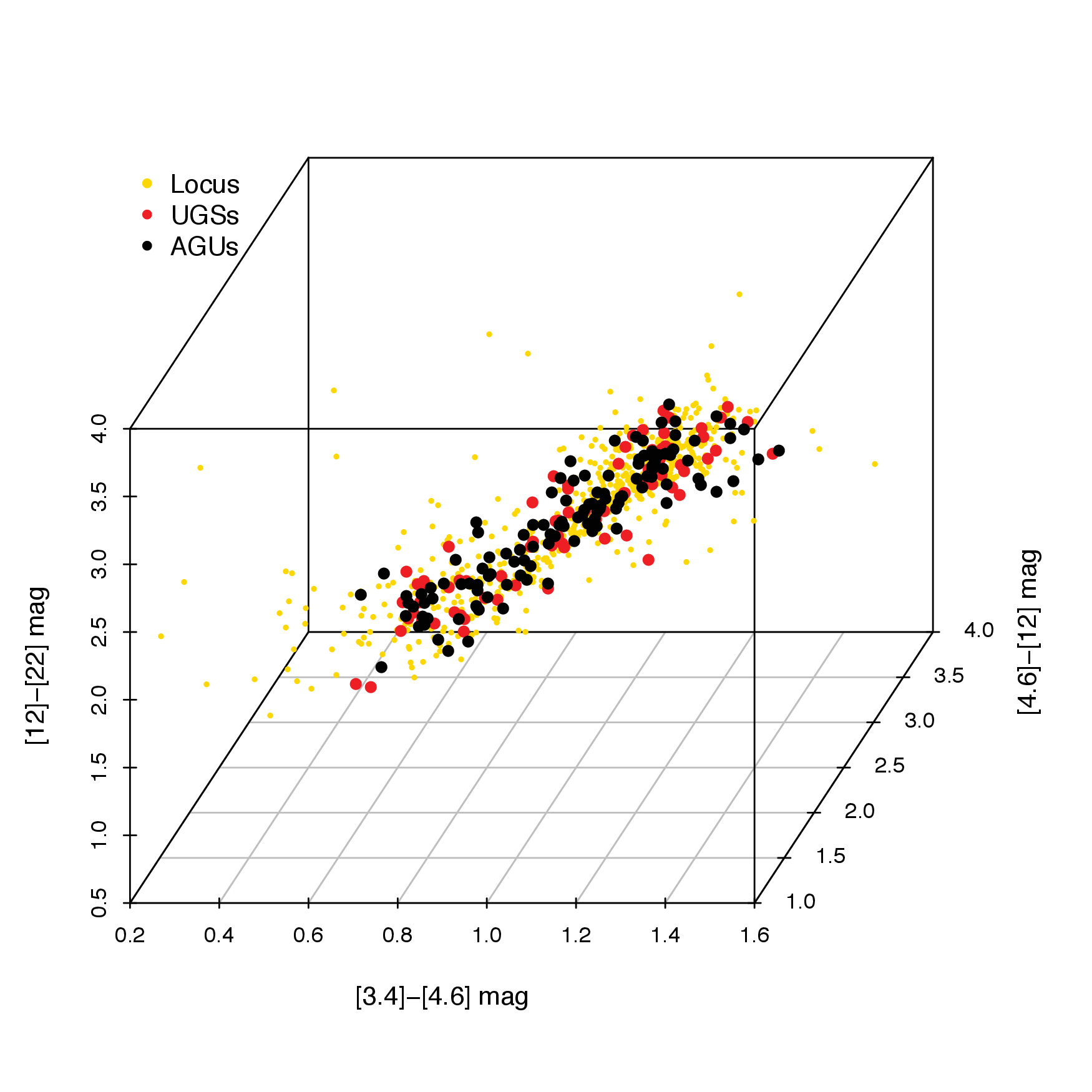}
          \caption{The 3D representation of the $locus$ (known $\gamma$-ray blazars are indicated in yellow)
                         in comparison with the selected $\gamma$-ray blazar candidates: UGSs (red) and AGUs (black).}
          \label{fig:wise}
          \end{figure}

We found 62 $\gamma$-ray blazar candidates for the UGSs without any $\gamma$-ray analysis flag
and 49 for those with $\gamma$-ray analysis flag, out of a total of 590 sources investigated.
For the AGU sample, we confirmed the blazar-like nature of 87 out 210 of AGUs 
analyzed on the basis of their IR colors.
Figure~\ref{fig:wise} shows the 3-dimensional color plot comparing the IR colors of the selected $\gamma$-ray blazar
candidates with the blazar population that constitutes the $locus$.

We searched in the following major radio,infrared, optical and X-ray surveys 
as well as in the NASA Extragalactic Database (NED)
\footnote{\underline{http://ned.ipac.caltech.edu/}}
for possible counterparts of our $\gamma$-ray blazar candidates, selected with the \wse\
association method, to verify if additional information could confirm their blazar-like nature. 
For the radio counterparts we used the NRAO VLA Sky Survey \citep[NVSS;][- N]{condon98}, 
the VLA Faint Images of the Radio Sky at Twenty-Centimeters \citep[FIRST;][- F]{becker95,white97}, 
the Sydney University Molonglo Sky Survey \citep[SUMSS;][- S]{mauch03}
and the Australia Telescope 20 GHz Survey \citep[AT20G;][- A]{murphy10}; 
for the infrared we used the Two Micron All Sky Survey \citep[2MASS;][- M]{skrutskie06}
since each \wse\ source is already associated with the closest 2MASS source 
by the default catalog \citep[see][for more details]{cutri12}.
Then, we also searched for optical counterparts, with possible spectra available, 
in the Sloan Digital Sky Survey \citep[SDSS; e.g.][- s]{adelman08,paris12}, in the Six-degree-Field Galaxy Redshift Survey 
\citep[6dFGS;][- 6]{jones04,jones09}; 
while for the high energy we looked in the soft X-rays using the ROSAT all-sky survey \citep[RASS;][- X]{voges99}.
We also searched in the USNO-B Catalog \citep{monet03} for the optical counterparts of our 
$\gamma$-ray blazar candidates

A significant fraction (i.e., $\sim$ 36\%) of the \wse\ sources associated with our method with UGSs
have a radio counterpart, more than 50\% are also detected in the 2MASS catalog as generally occurs
for blazars, and more than $\sim$10\% appear to be variable according to the \wse\ analysis flags \citep{cutri12}.
Notably, more than 90\% sources for which an optical spectrum was available in literature 
clearly show blazar-like features, being either featureless or having broad emission lines typical of quasars.
As generally expected for $\gamma$-ray blazars a handful of the selected candidates are also detected in the X-rays. 
A deeper investigation of their X-ray counterparts will be addressed in a forthcoming paper \citep{paggi13}.

Our results are in good agreement with those based on different statistical approaches
like the Classification Tree and the Logistic regression analyses \citep{ackermann12}.
In particular, 23 out of 28 UGSs that we associate to a $\gamma$-ray blazar candidate are also classified as active galaxies
by the above methods at high level of confidence.

Finally, we emphasize that additional investigations of different samples of active galactic nuclei, such as Seyfert galaxies, 
are necessary to study the problem of the contamination of our association method by 
extragalactic sources with infrared colors similar to those of $\gamma$-ray blazars. 
Moreover extensive ground-based spectroscopic follow up observations in the optical and in the near IR
would be ideal to verify the nature of the selected \wse\ sources and to estimate the fraction of non-blazar objects,
similar to the recent studies performed for the unidentified INTEGRAL sources \citep[e.g.,][]{masetti08,masetti09,masetti10,masetti12}.

\bigskip 
\begin{acknowledgments}
F. Massaro is grateful to S. Digel and D. Thompson for their helpful discussions
and to M. Ajello, E. Ferrara and J. Ballet for their support.
The work is supported by the NASA grants NNX12AO97G.
R. D'Abrusco gratefully acknowledges the financial 
support of the US Virtual Astronomical Observatory, which is sponsored by the
National Science Foundation and the National Aeronautics and Space Administration.
The work by G. Tosti is supported by the ASI/INAF contract I/005/12/0.
H. A. Smith acknowledges partial support from NASA/JPL grant RSA 1369566.
TOPCAT\footnote{\underline{http://www.star.bris.ac.uk/$\sim$mbt/topcat/}} 
\citep{taylor2005} and SAOImage DS9 were used extensively in this work 
for the preparation and manipulation of the tabular data and the images.
Part of this work is based on archival data, software or on-line services provided by the ASI Science Data Center.
This research has made use of data obtained from the High Energy Astrophysics Science Archive
Research Center (HEASARC) provided by NASA's Goddard
Space Flight Center; the SIMBAD database operated at CDS,
Strasbourg, France; the NASA/IPAC Extragalactic Database
(NED) operated by the Jet Propulsion Laboratory, California
Institute of Technology, under contract with the National Aeronautics and Space Administration.
Part of this work is based on the NVSS (NRAO VLA Sky Survey);
The National Radio Astronomy Observatory is operated by Associated Universities,
Inc., under contract with the National Science Foundation. 
This publication makes use of data products from the Two Micron All Sky Survey, which is a joint project of the University of 
Massachusetts and the Infrared Processing and Analysis Center/California Institute of Technology, funded by the National Aeronautics 
and Space Administration and the National Science Foundation.
This publication makes use of data products from the Wide-field Infrared Survey Explorer, 
which is a joint project of the University of California, Los Angeles, and 
the Jet Propulsion Laboratory/California Institute of Technology, 
funded by the National Aeronautics and Space Administration.
Funding for the SDSS and SDSS-II has been provided by the Alfred P. Sloan Foundation, 
the Participating Institutions, the National Science Foundation, the U.S. Department of Energy, 
the National Aeronautics and Space Administration, the Japanese Monbukagakusho, 
the Max Planck Society, and the Higher Education Funding Council for England. 
The SDSS Web Site is http://www.sdss.org/.
The SDSS is managed by the Astrophysical Research Consortium for the Participating Institutions. 
The Participating Institutions are the American Museum of Natural History, 
Astrophysical Institute Potsdam, University of Basel, University of Cambridge, 
Case Western Reserve University, University of Chicago, Drexel University, 
Fermilab, the Institute for Advanced Study, the Japan Participation Group, 
Johns Hopkins University, the Joint Institute for Nuclear Astrophysics, 
the Kavli Institute for Particle Astrophysics and Cosmology, the Korean Scientist Group, 
the Chinese Academy of Sciences (LAMOST), Los Alamos National Laboratory, 
the Max-Planck-Institute for Astronomy (MPIA), the Max-Planck-Institute for Astrophysics (MPA), 
New Mexico State University, Ohio State University, University of Pittsburgh, 
University of Portsmouth, Princeton University, the United States Naval Observatory, 
and the University of Washington.
\end{acknowledgments}

\bigskip 
{99}

\end{document}